\documentclass[ aip,jap, jmp,amsmath,amssymb,reprint]{revtex4-1}
\usepackage{graphicx}% Include figure files
\usepackage{dcolumn}% Align table columns on decimal point
\usepackage{bm}% bold math
\usepackage{natbib}
\begin{document}
\preprint{AIP/123-QED}
\title{Similar ultrafast dynamics of several dissimilar Dirac and Weyl semimetals} 
\date{\today}
\author{Chris P. Weber}\email{cweber@scu.edu}
\author{Bryan S. Berggren}
\author{Madison G. Masten}
\author{Thomas C. Ogloza}
\affiliation{Dept. of Physics, Santa Clara University, 500 El Camino Real, Santa Clara, CA 95053-0315, USA}
\author{Skylar Deckoff-Jones}
\author{Julien Mad\'{e}o}
\author{Michael K. L. Man}
\author{Keshav M. Dani}
\affiliation{Femtosecond Spectroscopy Unit, Okinawa Institute of Science and Technology Graduate University, 1919-1 Tancha, Onna-son, Kunigami, Okinawa 904-495, Japan}
\author{Lingxiao Zhao}
\author{Genfu Chen}
\affiliation{Institute of Physics and Beijing National Laboratory for Condensed Matter Physics, Chinese Academy of Sciences, Beijing 100190, China}
\author{Jinyu Liu}
\author{Zhiqiang Mao}
\affiliation{Department of Physics and Engineering Physics, Tulane University, New Orleans, Louisiana 70118, USA}
\author{Leslie M. Schoop}
\affiliation{Max Planck Institute for Solid State Research, Heisenbergstasse 1, 70569 Stuttgart, Germany}
\author{Bettina Lotsch}
\affiliation{Max Planck Institute for Solid State Research, Heisenbergstasse 1, 70569 Stuttgart, Germany}
\affiliation{Department of Chemistry, Ludwig-Maximilians-Universit\"at M\"unchen, Butenandtstrasse 5-13, 81377 M\"unchen, Germany}
\author{Stuart S. P. Parkin}
\author{Mazhar Ali}
\affiliation{Max Planck Institute of Microstructure Physics, Weinberg 2, 06120 Halle, Germany}

\begin{abstract}
Recent years have seen the rapid discovery of solids whose low-energy electrons have a massless, linear dispersion, such as Weyl, line-node, and Dirac semimetals. The remarkable optical properties predicted in these materials show their versatile potential for optoelectronic uses. However, little is known of their response in the picoseconds after absorbing a photon. Here we measure the ultrafast dynamics of four materials that share non-trivial band structure topology but that differ chemically, structurally, and in their low-energy band structures: ZrSiS, which hosts a Dirac line node and Dirac points; TaAs and NbP, which are Weyl semimetals; and Sr$_{1-y}$Mn$_{1-z}$Sb$_2$, in which Dirac fermions coexist with broken time-reversal symmetry. After photoexcitation by a short pulse, all four relax in two stages, first sub-picosecond, and then few-picosecond. Their rapid relaxation suggests that these and related materials may be suited for optical switches and fast infrared detectors. The complex change of refractive index shows that photoexcited carrier populations persist for a few picoseconds.
\\ \\
This article appeared in the \textit{Journal of Applied Physics}, \textbf{122}, 223102 (2017), and may be found at\\ \url{https://doi.org/10.1063/1.5006934}
\\
(This version of the article differs slightly from the published one.) 
\end{abstract}
\maketitle

\section{Introduction}

Interest has surged lately in topological, three-dimensional semimetals whose low-energy electron dispersions are linear and cross masslessly at a node. Several classes of such relativistic semimetals have been proposed, including line-node,\protect{\cite{Burkov2011LineNode}} Dirac,\protect{\cite{Young2012}} and Weyl.\protect{\cite{Wan2011,Burkov2011Weyl}} Subsequently Weyl nodes were predicted in a family of monopnictides,\protect{\cite{Weng2015,Huang2015NatCom}} and discovered in TaAs.\protect{\cite{Xu2015,Lv2015,Yang2015}} Beyond their fundamental interest, such materials could be technologically useful: they typically display high mobility and large magnetoresistance.\protect{\cite{Liang2014,Ali2014,HuangPRX2015,Shekhar2015,Ali2016}} It has been suggested that a $p$-$n$-$p$ junction of Weyl materials could act as a transistor despite the lack of an energy gap,\protect{\cite{Yesilyurt2016}} and that the materials could exhibit a large spin-Hall angle\protect{\cite{Sun2016}} and be ingredients in a quantum amplifier or a chiral battery.\protect{\cite{Kharzeev2013}} Their potential for optical and optoelectronic uses are enhanced by exotic predicted effects such as photocurrent driven by circularly-polarized mid-IR light,\protect{\cite{Chan2017}} anisotropic photoconductivity,\protect{\cite{Shao2015}} optical conductivity that takes the form of a step-function tunable by external fields,\protect{\cite{Ashby2014}} resonant transparency at THz frequencies tuned by a magnetic field\protect{\cite{Baum2015}}, and a mid-IR passband, tuned by the Fermi energy and lying between $E_F$ and $2E_F$.\protect{\cite{Kotov2016}}

\begin{figure*}
\centering
\includegraphics[width=5 in]{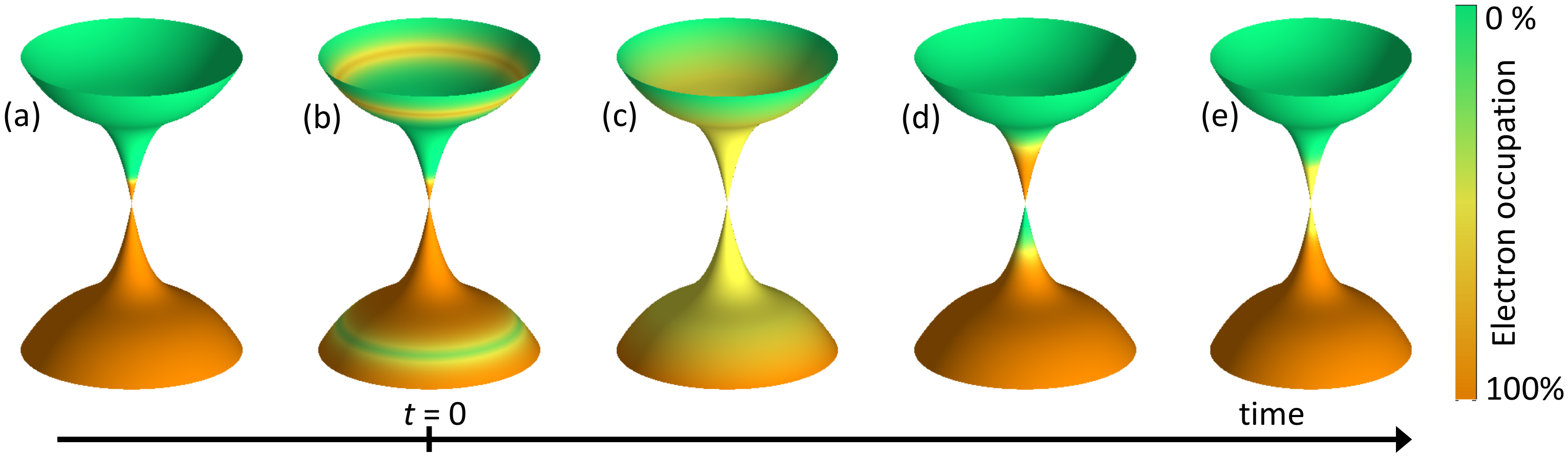}
\caption{\textbf{Schematic representation of the excitation and relaxation of electrons and holes in Cd$_3$As$_2$.} Shown as a function of time after photoexcitation, left to right. The many trivial bands far from the Fermi energy are represented by a single, broad continuum. (a) Prior to excitation, the material is slightly $n$-type. (b) Electrons and holes excited at high energy. (c) A high-temperature thermal distribution. (d) A partially-cooled distribution with inverted populations; it is unclear whether this situation occurs in Cd$_3$As$_2$. (e) A partially-cooled, non-inverted distribution; so long as the electronic temperature $T_e$ exceeds the lattice temperature, the carrier population remains thermally enhanced.}
\label{excitationdiagram}
\end{figure*}

Knowledge of a material's sub- and few-picosecond response to optical excitation---its ultrafast dynamics---holds practical significance. It reveals properties of the hot electrons important in high-field devices. More directly, it can guide optoelectronic applications: recently several devices have been reported that rely on the ultrafast properties of Cd$_3$As$_2$, the archetypal three-dimensional Dirac semimetal, to make fast photodetectors\protect{\cite{Wang2016,Yavarishad2017}} and optical switches.\protect{\cite{Zhu2017}} The monopnictide Weyl materials TaAs, TaP, and NbAs also show technological promise due to their sizable, anisotropic nonlinear-optical response.\protect{\cite{Wu2017}} Additionally, a broadband photodetector has recently been made out of TaAs.\protect{\cite{Chi2017}} The burgeoning variety of Dirac and Weyl semimetals, of diverse crystal and chemical structures, presents ever-wider opportunities for the materials' optoelectronic use; the need to explore and understand their ultrafast dynamics has grown commensurately. Knowledge of the materials' response to photoexcitation will likewise be important in realizing a predicted exciton condensate,\protect{\cite{Triola2017}} or various proposed effects in which intense pulses of light might separate or merge pairs of Weyl points or convert line nodes to point nodes.\protect{\cite{Ebihara2016,Chan2016,Narayan2016,Yan2016}} 

From recent ultrafast measurements on Cd$_3$As$_2$,\protect{\cite{Weber2015,Lu2016, Zhu2017b}} a picture is emerging in which its response to visible or near-infrared illumination is much like graphene's. Most electrons and holes are excited far from the Fermi energy, where the density of states is high (Fig. 1b). In tens or hundreds of femtoseconds, they share energy among themselves and with the resident charge-carriers to produce a quasi-thermal distribution (Fig. 1c) whose temperature, $T_e$, exceeds the lattice temperature; many of them occupy the Dirac cone. As the carriers cool (Fig. 1d-e), the narrowing of the Fermi-Dirac distribution further reduces the number of electrons and holes. In graphene, after forming the initial quasi-thermal distribution, electrons and holes may have separate chemical potentials\protect{\cite{Gilbertson2012,Gierz2013}}---that is, the carrier population may briefly be inverted (Fig. 1d)---recombining within a few hundred femtoseconds. Such inversion requires that the photocarriers' rate of cooling exceeds their recombination rate, and appears to be aided by the low density of states near the Fermi energy; indeed, it also occurs in the semimetals bismuth\protect{\cite{Sheu2013}} and graphite.\protect{\cite{Breusing2009}} In Cd$_3$As$_2$, however, it is not known whether a population inversion is ever formed.

Topological insulators (TIs) host surface states with a Dirac dispersion like that of the Dirac and Weyl semimetals. However, the ultrafast dynamics of these surface states are not closely analogous to those of the Dirac and Weyl semimetals, and instead are largely controlled by interactions between the bulk electronic states, which are gapped, and the surface.\protect{\cite{Sobota2012,Wang2012,Sim2014}} Optical excitation populates bulk states; a metastable population of these bulk electrons gradually feeds the population of the surface state; and the cooling of electrons in the surface state is strongly influenced by bulk-surface coupling.

In three-dimensional Dirac semimetals other than Cd$_3$As$_2$, time- and angle-resolved photoemission experiments support the same picture as in Cd$_3$As$_2$. In the Weyl material MoTe$_2$,\protect{\cite{Wan2017}} and the gapped Dirac materials ZrTe$_5$\protect{\cite{Manzoni2015}} and SrMnBi$_2$,\protect{\cite{Ishida2016}} electrons excited by 1.5-eV photons relax into the Dirac cone within about 0.4 ps. The final stage of electronic cooling in SrMnBi$_2$ slows to a power-law,\protect{\cite{Ishida2016}} consistent with predictions for Dirac and Weyl semimetals.\protect{\cite{Lundgren2015,Bhargavi2016}} In MoTe$_2$, the electrons' non-equilibrium temperature was seen to recover as a biexponential, with time constants 0.43 ps and 4.1 ps; in contrast to graphene, the population inversion illustrated in Fig. 1d was not observed.\protect{\cite{Wan2017}}

\begin{figure*}
\centering
\includegraphics[width=\textwidth]{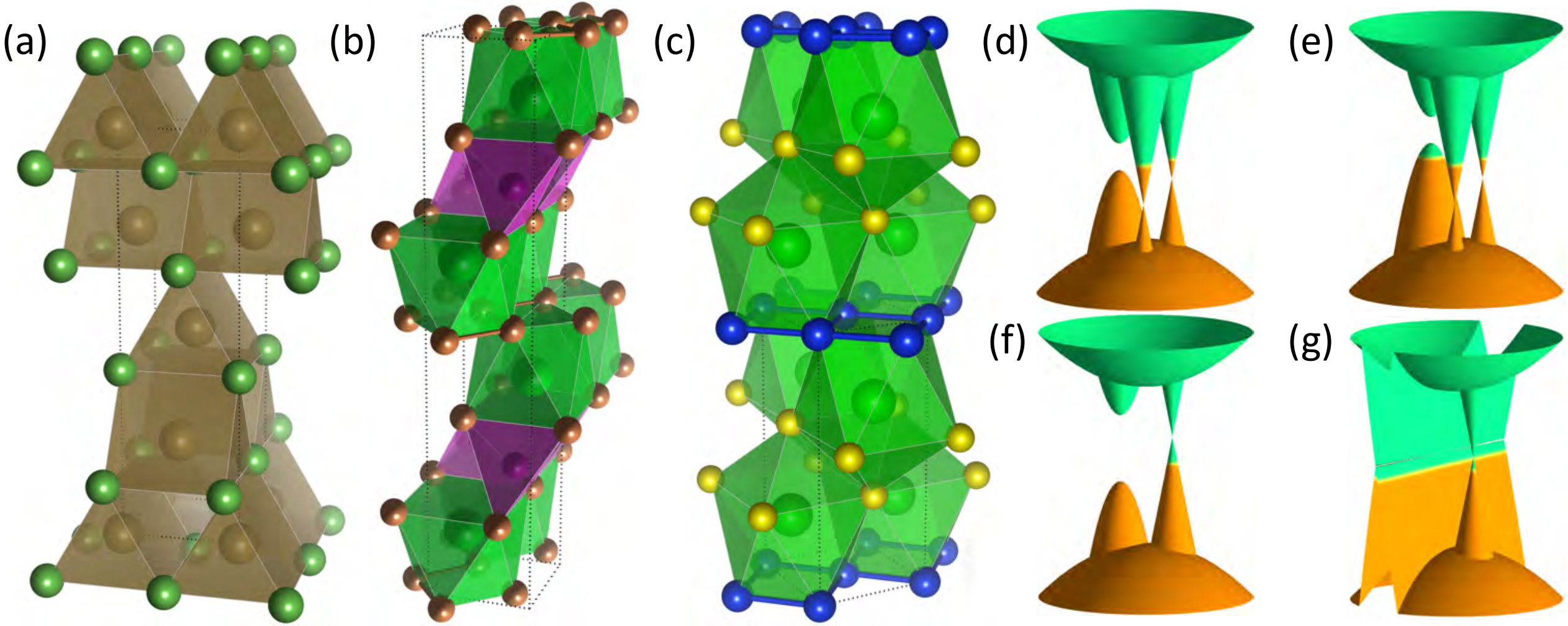}
\caption{\textbf{Crystal and band structures. (a)-(c)}: Crystal structures. \textbf{(a)} TaAs and NbP. Shown are the As$_6$ (or P$_6$) prisms, with a single Ta (or Nb) at the center of each. \textbf{(b)} SrMnSb$_2$, with Sb atoms shown in brown. The Sb$_4$ tetrahedra (magenta) each contain a Mn atom, while the Sb$_8$ square antiprisms (green) each contain a Sr atom. \textbf{(c)} ZrSiS, with Si in blue and S in yellow. Each mono-capped square antiprism consists of a Zr atom coordinated by four Si atoms forming a planar square net and five S atoms. \textbf{(d)-(g)}: Schematic low-energy band structures. The many trivial bands far from the Fermi energy are represented by a single, broad continuum. Features that occur at many points in the Brillouin zone are shown just once; nor is chirality indicated. Filled and empty states are indicated in the same color-scale as Fig. 1. \textbf{(d)} In TaAs, there are $n$-type linear cones of two different energies. \textbf{(e)} NbP has a band structure like that of TaAs, but a different Fermi level introduces $p$-type massive carriers. \textbf{(f)} In Sr$_{1-y}$Mn$_{1-z}$Sb$_2$ transport is dominated by a $p$-type linear cone. \textbf{(g)} ZrSiS, showing a line node  and a Dirac cone (both shown $p$-type).}
\label{structureanddispersion}
\end{figure*}

However, the most widespread measure of the ultrafast response, and that employed in this work, is the pump-induced change in reflectivity, $\Delta R(t)$, of a time-delayed probe pulse. While photoemission measures carrier populations near the Fermi energy, reflectivity measures changes in the index of refraction at the probe energy. This energy is well beyond the Dirac cone, but through the Kramers-Kronig relation $\Delta R(t)$ measures carrier populations at many energies. Indeed, work on Cd$_3$As$_2$ has shown that $\Delta R(t)$ primarily reveals the lifetime of the carriers near the Dirac point: the timescale of the ultrafast response was nearly independent of the probe photons' energy, but its magnitude increased as the probe's energy was lowered toward the Fermi level.\protect{\cite{Lu2016}} The materials' optoelectronic properties thus persist from the visible through the mid-infrared.\protect{\cite{Zhu2017,Wang2016}}

In this work we investigate the ultrafast dynamics of four Dirac and Weyl materials: TaAs and NbP;\protect{\cite{Weng2015,HuangPRX2015}} ZrSiS;\protect{\cite{Schoop2016}} and Sr$_{1-y}$Mn$_{1-z}$Sb$_2$.\protect{\cite{Liu2015}} As shown in Fig. 2, these materials differ sharply in chemical and crystal structure. TaAs and NbP crystallize in space group \textit{I}$4_1$\textit{md}, (\#109) which can be thought of as a network of face- and edge-sharing TaAs$_6$ (or NbP$_6$) trigonal prisms. SrMnSb$_2$ (\textit{Pnma}, \#62) has layers of edge-sharing MnSb$_4$ tetrahedra spaced by a bilayer of face-sharing SrSb$_8$ square antiprisms. In the shared Sb layer, the Sb atoms are distorted from square nets to be slightly orthorhombic and to form zig-zag chains. This contrasts with ZrSiS (\textit{P}4/\textit{nmm} \#129), which can be thought of as layers of mono-capped square antiprisms with Zr coordinated by Si and S with the Si atoms forming a planar square net between Zr layers. The materials' one common feature is that their electronic structures, near $E_F$, have linear band crossings and a local minimum in the density of states---though even these differ in origin and type. ZrSiS is a Dirac line-node semimetal and, by virtue of having both inversion and time-reversal symmetry, has Dirac nodes near $E_F$ that are gapped to a small extent by spin-orbit coupling.\protect{\cite{Schoop2016}} TaAs and NbP are inversion-breaking Weyl semimetals, resulting in doubly degenerate Weyl points near $E_F$,\protect{\cite{Weng2015,Huang2015NatCom}} and negative magnetoresistance induced by the chiral anomaly has been observed in TaAs.\protect{\cite{HuangPRX2015}} Sr$_{1-y}$Mn$_{1-z}$Sb$_2$ preserves inversion symmetry but breaks time-reversal symmetry due to magnetic ordering. The Sb plane gives rise to nearly massless Dirac fermions.\protect{\cite{Liu2015}} 

As our experiments reveal, these four dissimilar materials share very similar ultrafast responses, consisting of sub-picosecond and few-picosecond components. Moreover, the results of our phase-sensitive transient-grating measurements indicate that a carrier population lasts for at least 1ps. These four materials' responses are similar to that of Cd$_3$As$_2$, but even faster, suggesting that they, too, may be well-suited to optical switches and photodetectors. As the discovery of additional Dirac and Weyl semimetals proceeds, other materials are likely to follow the same pattern---and therefore to enable a wide range of optoelectronic applications.

\section{Methods}
\begin{figure*}
\centering
\includegraphics[width=\textwidth]{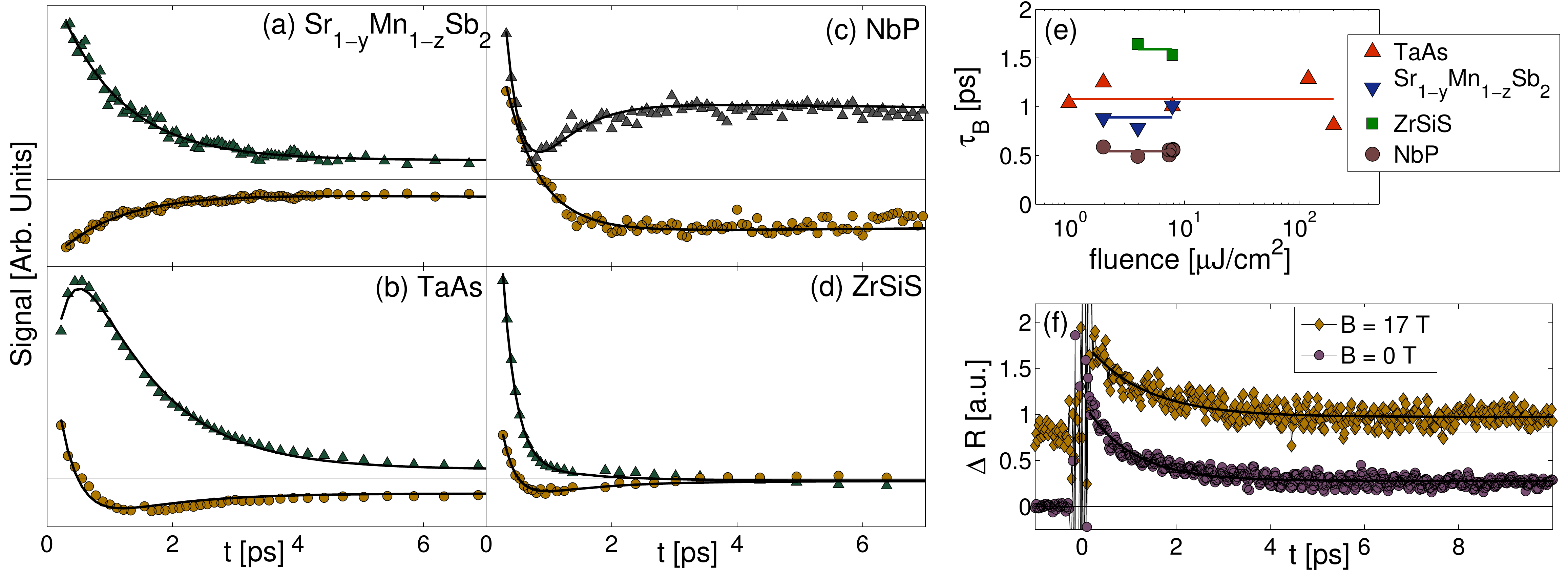}
\caption{\textbf{Ultrafast responses. (a-d)}: Typical transient-grating data at room temperature for the four materials studied. Triangles and circles are real and imaginary parts, respectively (up to a factor of $e^{i\phi}$). Solid curves are fits to Eq. \ref{twoexp}. \textbf{(e)}: Fluence-dependence of the $\tau_B$ decay at room temperature. The two highest-fluence points for TaAs are measured with the probe energy below the pump; see Fig. 4a. \textbf{(f)}: Sr$_{1-y}$Mn$_{1-z}$Sb$_2$: 0 T (circles), 17 T (diamonds), both 10 K. Shifted for clarity. What looks like scatter in the data is actually oscillations. The smooth curves are fits to a single exponential plus an offset, and give $\tau_B$ of 1 ps and 1.3 ps, respectively. This experiment was pump-probe but not transient-grating, and thus measured $\Delta R(t)$, which is real-valued.}
\label{ppstacked}
\end{figure*}

The single crystals of TaAs were grown by the chemical vapor transport method which was described in Ref. \protect{\onlinecite{HuangPRX2015}}. Samples grown in this way show electronic mobility reaching $1.8\times10^5$ cm$^2/$Vs at 1.8 K. The samples of Sr$_{1-y}$Mn$_{1-z}$Sb$_2$ were grown using a self-flux method as detailed in Ref. \protect{\onlinecite{Liu2015}}. They were ``Type B,'' in the terminology of Ref. \protect{\onlinecite{Liu2015}}, with $y\approx0.06$, $z\approx0.08$, about 0.08 $\mu_{\text{B}}$ per Mn and a mobility near $10^4$ cm$^2$/Vs. SdH oscillations on a sample from this batch had a frequency in the range of $66-70$ T, corresponding to a very small Fermi surface with $A_F=0.64$ nm$^{-2}$. Transport is hole-dominated. This type of sample harbors nearly massless Dirac fermions. Single crystals of ZrSiS and NbP were both grown via iodine vapor transport. For ZrSiS, the synthesis followed the method of Ref. \protect{\onlinecite{Schoop2016}}. For NbP, single crystals were grown from a polycrystalline powder of NbP (obtained from direct reaction of Nb and P at 800 $^\circ$C for five days) at 950 $^\circ$C with a temperature gradient of 100 $^\circ$C. The crystals were obtained towards the hot end of the tube and then annealed at 500 $^\circ$C for one week.

All ultrafast measurements used mode-locked Ti:Sapphire lasers (or, in Fig. 3f, a fiber laser). For the measurements at high fluence and wavelengths other than 800 nm (Figs. 3e and 4a), a long-cavity laser and optical parametric amplifier were used, and the beams were focused on the sample through a microscope objective. In Fig. 4d, we determine the complex phase of $\Delta n(t)$ from a pair of transient-grating measurements by the method described in Refs. \protect{\cite{Gedik2004,Weber2015}}, and in the Supplementary Materials section S1, and using calculations of the static index $n$ found in Ref. \protect{\onlinecite{Buckeridge2016}}. All measurements used standard chopping and lock-in detection.

\section{Results}

\subsection{Ultrafast response}
After a sample absorbs a short pulse of light, the ``pump,'' we probe the dynamics of the photoexcited electrons by measuring the pump-induced change in the reflectivity, $\Delta R(t)$, of a ``probe'' pulse that arrives a time $t$ after the pump. We also perform transient-grating measurements, an extension of the pump-probe method in which the diffraction of the probe is measured. For our purposes, the transient grating's significance lies primarily in improved signal. It also is phase-sensitive, measuring the real and imaginary parts of the change in reflectance, $\Delta r(t)$, where $R=rr^*$. While $\Delta r(t)$ is typically known only up to an overall complex phase of $e^{i\phi}$, we will show (in Fig. 4d) a measurement of $\phi$ that allows us to determine the change in the index of refraction, $\Delta n(t)$.

\begin{table*}\
\caption{\textbf{Room-temperature decay rates} for the materials studied. Rapid oscillations in the signal of Sr$_{1-y}$Mn$_{1-z}$Sb$_2$ make it impossible to determine $\tau_A$.} 
\vspace*{1.5\baselineskip}
\begin{tabular}{l|c|c|c|c|c}
\hline Material & Sr$_{1-y}$Mn$_{1-z}$Sb$_2$ & TaAs & NbP & ZrSiS & Cd$_3$As$_2$ [Ref. \protect{\onlinecite{Weber2015}}] \\ \hline\hline
$\tau_A [ps]$ & $N/A$ & $0.38\pm0.13$ & $0.27\pm0.06$ &  $0.19\pm0.03$ & $0.50\pm0.04$  \\ \hline
$\tau_B [ps]$ & $0.96\pm0.18$ & $1.1\pm0.1$ & $0.50\pm0.08$ &  $1.6\pm0.3$ & $3.1\pm0.1$ \\ \hline
\end{tabular}
\label{tabletau}
\end{table*}

Examples of the transient-grating data appear in Fig. 3. Each material's $\Delta r(t)$ appears to have a different shape, but the differences lie mostly in the arbitrary phase $\phi$ and in the size of a nearly constant component that represents heating of the lattice (discussed below; also see Supplementary Materials section S2, Fig. S3, and Table S1). For instance, the real part of the NbP signal first dips then rises; but for a different phase $\phi$ it would first rise then fall, resembling TaAs. 

Despite the differences of shape, it is the similarities in the materials' dynamics that are much more striking---along with their similarities to prior measurements of Cd$_3$As$_2$.\protect{\cite{Weber2015}} For each material, the ultrafast response on the timescale of interest fits well to a biexponential plus a constant:
\begin{equation}\Delta r(t)=Ae^{i\theta_A}e^{-t/\tau_A}+Be^{i\theta_B}e^{-t/\tau_B}+Ce^{i\theta_C},
\label{twoexp}\end{equation}
with $\tau_A$ sub-picosecond and $\tau_B$ several times longer (see Table 1). The slower decay has an amplitude typically about $20\%$ that of the fast one. (We found that attempts to fit the data to simpler functions such as a single exponential or a bimolecular decay were unsuccessful; see Supplementary Materials section S3 and Fig. S4.)

Another similarity among these materials, and a similarity to Cd$_3$As$_2$, is that the ultrafast response is nearly independent of many experimental conditions. For all four materials, we see no difference between measurements at room temperature and at 10 K. For TaAs, we also measured temperatures from 80 to 230 K, a range over which the carrier densities $n$ and $p$ change by factors of 20;\protect{\cite{Zhang2015}} nonetheless, $\tau_A$ and $\tau_B$ remained constant. The fluence, or energy per area, of the pump pulse also does not change the decay rate (Fig. 3e). For Sr$_{1-y}$Mn$_{1-z}$Sb$_2$ we measured up to $B=17$ T, with little or no change in the signal (Fig. 3f); magnetic field likewise had no effect on the responses of TaAs (see Supplementary Materials section S4 and Fig. S5) or ZrSiS, though measured only up to $B=0.3$ T. Naturally, the materials' responses do have some differences: Cd$_3$As$_2$ is the slowest, and NbP is the fastest; TaAs has a larger amplitude ratio $B/A$; NbP has the largest lattice-heating term. For Sr$_{1-y}$Mn$_{1-z}$Sb$_2$ the signal includes oscillations at a few THz, which overlap with the timescale of the fast decay and prevent us from determining $\tau_A$. (These oscillations will be discussed in a separate publication.) Nonetheless, the very fast, two-part responses are notably similar for such dissimilar materials, contrasting both with other types of semimetal and with metals, as we will discuss below. This similarity raises the hope that such a sub-picosecond and few-picosecond response may be generic to Dirac and Weyl materials, including those yet to be explored. 

\subsection{Physical origins of the ultrafast response}

A full accounting of the many processes leading to the relaxation on timescales $\tau_A$ and $\tau_B$ in these four materials would require many more experiments. However, the broad outlines of the processes are apparent. 

First, consider the excitation of electrons and holes by absorption of pump photons of energy 1.55 eV, an energy that exceeds the extent of the Dirac or Weyl cones. The dominant optical transitions will be those with the largest joint density of states, and in semimetals the density of states may be several orders of magnitude smaller near $E_F$ than at higher and lower energies.\protect{\cite{Sheu2013}} The pump therefore initially excites carriers into the massive bands beyond the range of the linear dispersion, as illustrated in Fig. 1b. 

The fast process we observe, $\tau_A$, most likely arises from the scattering of electrons and holes out of these initial high-energy states. The electrons and holes form a ``hot'' distribution spread over a broad energy range and have temperatures that greatly exceed the lattice temperature (Fig. 1c). The scattering may be both intraband and interband, and may include electron-electron and electron-phonon processes, both of which are hastened by the large density of states far from $E_F$. Additionally, electron-electron scattering is faster in semimetals than in metals because the Coulomb interaction is less screened. 

\begin{figure*}
\includegraphics[width=\linewidth]{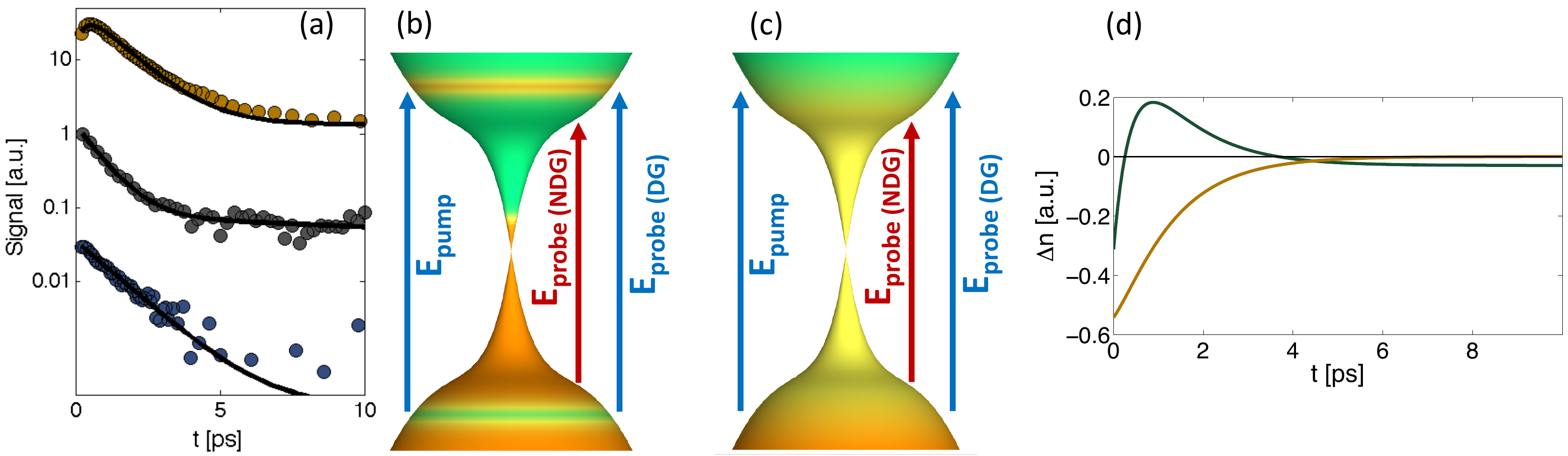}
\caption{\textbf{Data for TaAs at room temperature.} (a) Ultrafast response for various pump and probe energies. Top: pump 1.55 eV, probe 1.55 eV. Middle: pump 1.55 eV, probe 1.10 eV. Bottom: pump 2.36 eV, probe 1.55 eV. The curves are shifted vertically for clarity. The data are $\Delta R(t)$, which is real-valued. (b,c) Schematic illustration of the photon energies used as pump (blue) and probe in degenerate (DG) and non-degenerate (NDG) experiments, along with the electron occupation before (b) and after (c) interband scattering. (d) The photoinduced change in refractive index, $\Delta n(t)$; the real part is green, and the imaginary part is orange. The curve is calculated from the measured phase of $\Delta r(t)$ and from the data's fit to Eq. \ref{twoexp}, as explained in Ref. \protect{\onlinecite{Weber2015}}. }
\label{LambdaAndPhase}
\end{figure*}

Most of our experiments are ``degenerate,'' with the pump and probe photons having the same energy of 1.55 eV. To confirm that most carriers lie below the pump energy within a picosecond, we did non-degenerate measurements on TaAs with a 1.55-eV pump and a 1.10-eV probe, and again with a 2.36-eV pump and a 1.55-eV probe. The measurement did not resolve $\tau_A$, but $\tau_B$ remained the same as in the degenerate experiments, as shown in Fig. 4a. This independence of probe wavelength would seem very implausible if the carriers still resided largely at the pump energy, as in Fig. 4b; in that case, their strong filling of phase space would cause a probe at the pump energy to measure starkly different dynamics than one at lower energy. On the other hand, after electrons have scattered to lower energy (Fig. 4c), the degenerate and non-degenerate probes would yield similar results. Lu \textit{et al.}\protect{\cite{Lu2016}} obtained a similar result, with a similar interpretation, in Cd$_3$As$_2$ as they lowered the probe energy closer to the Fermi surface.

As these carriers scatter to lower energy, many end up in the Dirac or Weyl cones---as, indeed, they are known to do within a picosecond in ZrTe$_5$\protect{\cite{Manzoni2015}} and SrMnBi$_2$.\protect{\cite{Ishida2016}} They subsequently persist at low energies for a few picoseconds, giving rise to the signal $\Delta r(t)$ on the timescale of $\tau_B$. Their persistence is not obvious \textit{a priori}, but we support it by measuring the complex phase of our signal, $\Delta r(t)$. We express the result in terms of the change in the complex index of refraction, $\Delta n(t)$, where $n=n_r+in_i$. To understand $\Delta n(t)$, recall that the measured reflectance depends on the refractive index evaluated at the probe's frequency; $n_i$ represents absorption, and $n_r$ refraction. When photoexcitation changes the absorption at \textit{any} frequency, there is a resulting change $\Delta n_r$ at the probe frequency due to the Kramers-Kronig relations between $n_r$ and $n_i$.\protect{\cite{Hutchings1992}} 

The results for TaAs appear in Fig. 4d, and a similar result for ZrSiS appears in the Supplementary Materials, Section S1 and Figs. S1 and S2. The key observation is that from 0.25 ps to 3.6 ps $\Delta n_r$ is positive, while $\Delta n_i$ is negative. These signs are the signature of phase-space filling (PSF)---the occupation of states by photoexcited electrons and holes---at energies below the probe energy.\protect{\cite{Hutchings1992}} PSF reduces the optical absorption at the corresponding energy; since the carriers have a broad thermal distribution, PSF extends even to the probe energy (Fig. 1c), causing the observed $\Delta n_i<0$. The corresponding Kramers-Kronig change to $n_r$ is positive. It is noteworthy that a negative $\Delta n_i$ and a positive $\Delta n_r$ was also observed in Cd$_3$As$_2$ on a similar time-scale.\protect{\cite{Weber2015}} 

After a few picoseconds, the phase of $\Delta n(t)$ changes. At all later times, $\Delta n_i>0$ and $\Delta n_r<0$. These signs are characteristic of the other two primary mechanisms by which photoexcitation changes the optical absorption, namely Drude absorption and band-gap renormalization (BGR);\protect{\cite{Hutchings1992}} our experiment does not distinguish the two mechanisms. In Drude absorption, the photoexcited electrons and holes increase the free-carrier density, and the optical conductivity, at low energies. In BGR, heating of the lattice reduces the energy-gap between valence and conduction bands. (In Dirac and Weyl materials, though the linearly-dispersing bands lack a gap, BGR may influence the gap between the higher-energy, massive bands.) The slowest component of our signal lasts for hundreds of picoseconds, suggesting that it arises from lattice heating and BGR. Moreover, in ZrSiS we were able to measure the decay rate and diffusion coefficient of this signal (see Supplementary Materials section S2 and Fig. S3), which are consistent with thermal transport.

\section{Discussion}

We now relate the observed ultrafast response to the low-energy electronic structure of Dirac and Weyl semimetals, particularly to the low density of states near $E_F$ and the linear dispersion. During the few-picosecond $\tau_B$ response, photoexcited electrons and holes are recombining. The low density of states slows the recombination considerably by restricting the phase-space for scattering between electron and hole states. In metals, by contrast, recombination is a purely intraband process, and proceeds at the much faster rate of electron-electron scattering,\protect{\cite{Hohlfeld2000,DelFatti2000}} roughly equivalent to our sub-picosecond $\tau_A$; recombination may be followed by a few-picosecond phase of electronic cooling. The similarity of metals' ultrafast response to those of Dirac and Weyl semimetals thus belies its very different physical origin. Metals, moreover, lack the strong nonlinearities seen in Dirac and Weyl semimetals,\protect{\cite{Wu2017}} limiting their optoelectronic uses. 

Though photocarriers in our Dirac and Weyl materials last much longer than in metals, they last a much shorter time than in some other semimetals. In bismuth---a semimetal with a momentum-space gap---recombination requires the assistance of a high-momentum phonon, and accordingly lasts for 12 to 26 ps.\protect{\cite{Sheu2013}}  WTe$_2$, a type-II Weyl semimetal, is similar: in samples with a momentum gap at $E_F$, $\Delta R(t)$ decays in two parts, with $\tau_B=5$ to 15 ps.\protect{\cite{Dai2015,He2016}} In contrast, the linear, Dirac-like dispersion in our materials speeds recombination because it enables low-momentum transitions between electron and hole states. The linear dispersion also ensures that Auger processes (both interband and intraband) automatically satisfy both energy and momentum conservation, provided they occur along a straight line in momentum-space. Similar considerations make Auger recombination very efficient in graphene.\protect{\cite{Winzer2010}}

A third key characteristic of Dirac and Weyl semimetals, their lack of an energy gap at the node, is less apparent in the ultrafast response. Opening a gap can shorten the electronic lifetime by increasing the density of states for electron-phonon scattering,\protect{\cite{Narang2016}} but can also lengthen the lifetime by reducing the phase-space available to Auger recombination, as seen in bilayer graphene \protect{\cite{Gierz2015}}. Moreover, a nonzero $E_F$ would have much the same effect on the ultrafast response as would a gap: it would slow Auger recombination by restricting its phase-space.

Though we have focused on our materials' similar ultrafast responses, their decay rates do differ by factors of 2 to 3. A number of material-dependent effects can influence these decay rates, including screening by bound electrons,\protect{\cite{DelFatti2000}} the phonon band-structure,\protect{\cite{Lundgren2015}} and the size of the Fermi surface. The latter affects screening of electron-electron interactions, the phase-space available to Auger processes, and plasmon emission.

As new Dirac and Weyl semimetals are discovered, whether in the same material families as our samples or in new ones, they will all share the linear dispersion and low density of states near $E_F$ that dictate the ultrafast properties of our samples. Thus we can anticipate a rapid, two-part ultrafast response, similar to the one which makes the four materials we've studied promising for optoelectronic applications such as terahertz detectors or saturable absorbers. This ultrafast response could be especially useful if combined with the materials' demonstrated large, anisotropic optical nonlinearity,\protect{\cite{Wu2017}} or with any of the materials' remarkable predicted optical properties.\protect{\cite{Chan2017,Shao2015,Ashby2014,Baum2015,Kotov2016}}

\section*{Supplementary Material}

See supplementary material for: discussion of the signal's complex phase; discussion of the signal at long times; examples of fits of the data to functional forms other than Eq. \ref{twoexp}; and additional data taken in a magnetic field.

\section*{Acknowledgements}
Work in Santa Clara and Okinawa was supported by the National Science Foundation DMR-1508278 and by the Geoff and Josie Fox Scholarship. Work in Beijing was supported by the National Basic Research Program of China 973 Program (Grant No. 2015CB921303), the National Key Research Program of China (Grant No. 2016YFA0300604), and the Strategic Priority Research Program (B) of Chinese Academy of Sciences (Grant No. XDB07020100). Work in New Orleans was supported by the U.S. Department of Energy under EPSCoR Grant No. DESC0012432 with additional support from the Louisiana Board of Regents. Work in Germany was supported by the Max Plank Institute (MPI) for Microstructure Physics in Halle, the MPI for Solid State Research in Stuttgart, and the Alexander von Humboldt Foundation.

\renewcommand\refname{References and Notes}

\renewcommand\thefigure{S\arabic{figure}} 
\renewcommand\theequation{S\arabic{equation}} 
\renewcommand{\thesection}{S\arabic{section}}  
\renewcommand{\thesubsection}{S\arabic{section}.\arabic{subsection}}  
\renewcommand{\thetable}{S\arabic{table}} 

\section*{Supplementary material}
\setcounter{section}{0}
\setcounter{figure}{0}
\setcounter{table}{0}
\section{Absolute phase of transient-grating signal}
\subsection{Complex $\Delta n$ in ZrSiS}

\begin{figure}
\includegraphics[width=3.37 in]{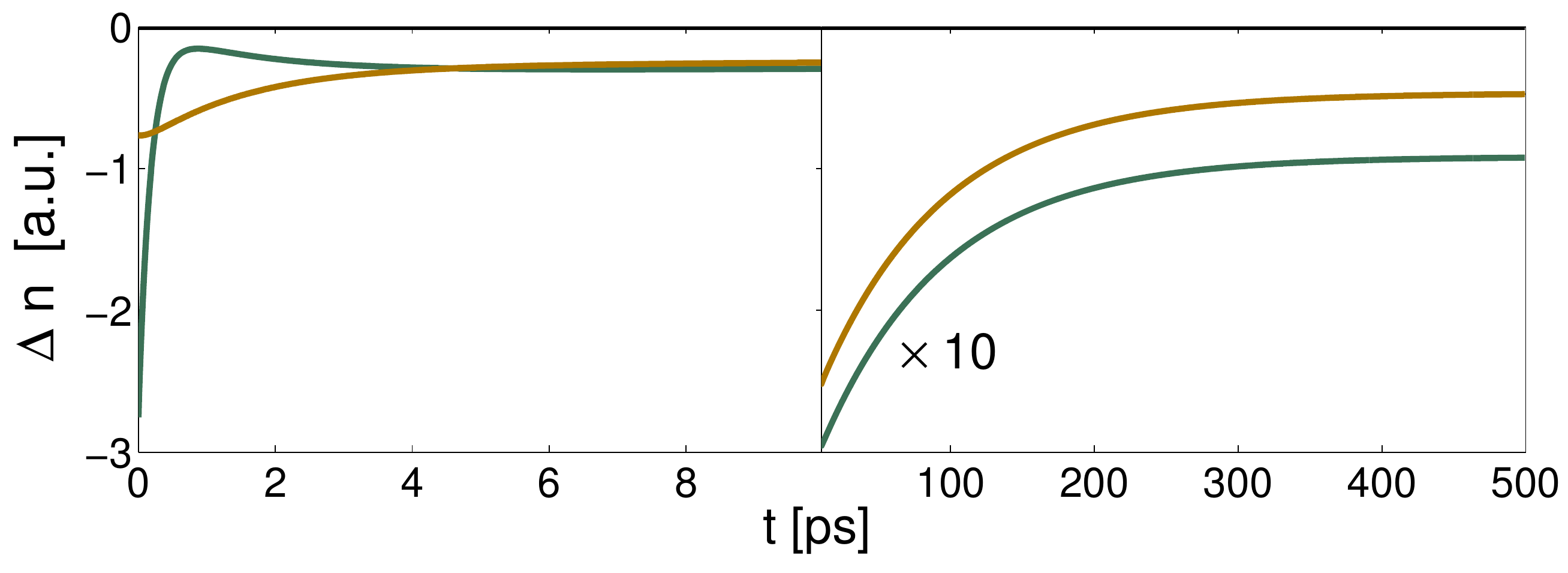}
\caption{\textbf{The photoinduced change in refractive index, $\Delta n(t)$, for ZrSiS at room temperature.} The real part is green, and the imaginary part is orange. The curve is calculated from the measured phase of $\Delta r(t)$ and from the data's fit to Eq. \ref{twoexpsup}, as explained in Ref. \protect{\cite{Weber2015}}.}
\label{Delta_n_ZrSiS}
\end{figure}

In the main text, we showed, for TaAs, $\Delta n(t)$---that is, the photoinduced change in the refractive index. In Fig. \ref{Delta_n_ZrSiS}, we show $\Delta n(t)$ for ZrSiS. It is immediately apparent that the results for TaAs and ZriSiS are very similar. The primary difference appears in the real part $\Delta n_r(t)$ for times 1 ps $<t<$ 3 ps. At those times in TaAs, the real part $\Delta n_r(t)>0$, while in ZrSiS $\Delta n_r(t)$ also rises to a local maximum, but it never quite exceeds zero. We interpret this difference as arising from the larger contribution of the ``$C$'' component in ZrSiS---that is, of the long-time thermal signal. This signal (shown expanded in the right panel of Fig. \ref{Delta_n_ZrSiS}) has the effect of making both $\Delta n_r$ and $\Delta n_i$ more negative. Nonetheless, the result in ZrSiS generally agrees with that in TaAs.

\subsection{Static index of refraction of ZrSiS}

In determining the photoinduced change of refractive index, $\Delta n$, it was necessary to know the unperturbed index, $n=n_r+in_i$. For ZrSiS no values of $n$ have been published, so we estimated it by measuring the sample's reflectivity for both s- and p-polarized light as a function of  the incident angle. (These measurements used the same 800-nm laser as our transient-grating measurements.) We then determined $n$ by fitting the data to the Fresnel equations. Data and fits appear in Fig. \ref{Delta_n_ZrSiS}.

\begin{figure}
\includegraphics[width=2.85 in]{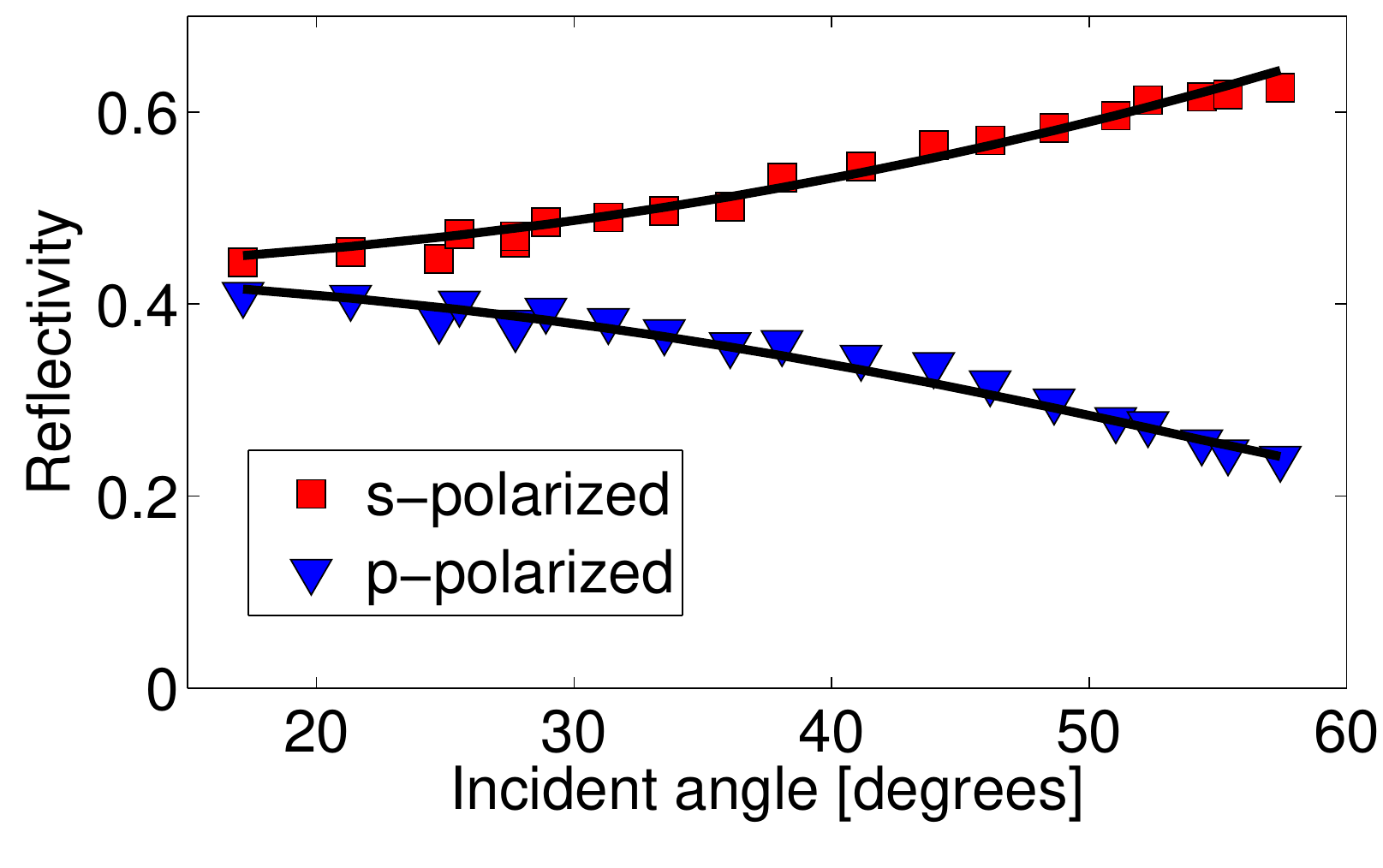}
\caption{\textbf{Reflectivity of ZrSiS \textit{vs}. incident angle.} At 800 nm. Lines are fits to the data using the Fresnel equations, with the complex index $n$ as a free parameter.}
\label{reflectivityZrSiS}
\end{figure}

For $n_i=0$, it is well known that at Brewster's angle the reflectivity of p-polarized light, $R_p(\theta)$, is zero. When $n_i\neq0$, there is still a minimum in $R_p(\theta)$ near Brewster's angle, but the reflectivity does not become zero. We were prevented from measuring to large incident angles---and to Brewster's angle---by geometric constraints of our setup and by the size of the sample. Though such measurements would have improved the accuracy of $n_i$, the fits shown in Fig. \ref{reflectivityZrSiS} gave just modest error bars: $n_r=1.8\pm0.3$ and $n_i=2.21\pm.16$. The combined uncertainty in the real and imaginary parts leads to an uncertainty of 24 degrees in the polar angle of $n$.

To validate this experimental method, we repeated the measurement of reflectivity \textit{vs}. angle on a TaAs sample. We obtained $n_r$ and $n_i$ in good agreement with published values.\protect{\cite{Buckeridge2016}

\subsection{Measurement and calculation of absolute phase}

In a heterodyne-detected transient-grating experiment, it is possible to determine the phase of the diffracted signal\protect{\cite{Gedik2004, Weber2015} by comparing four measurements: of the beam diffracted in the $m=+1$ or $m=-1$ order; for each of these, the real (imaginary) part of our signal corresponds to measurement without (with) an additional phase of $\pi/2$ introduced between the probe and the local oscillator. Thus we obtain two complex diffracted signals, $d^{+1}(t)$ and $d^{-1}(t)$, which are fit to:
\begin{equation}d^{\pm1}(t)=Ae^{i\theta^{\pm1}_A}e^{-t/\tau_A}+Be^{i\theta^{\pm1}_B}e^{-t/\tau_B}+Ce^{i\theta^{\pm1}_C}
\end{equation} 
From these measured values of $\theta^{\pm1}$, there are several steps to determining the complex $\Delta n(t)$. We define the sample's reflectance $r$ as $r=|r|e^{i\phi_r}$, and determine $\phi_r$ from Fresnel's equations and the refractive index $n$ (in the absence of photoexcitation), which is known or measured. We can then calculate
\begin{equation}
\phi_{\Delta r}=\frac{\theta^{-1}-\theta^{+1}-\pi}{2}+\phi_r,
\label{halfangle}\end{equation}

We note that the photoinduced change in reflectance is related to the change in index by
\begin{equation}\Delta n=-\left[\frac{(1+n)^2}{2}\right]\Delta r,
\label{Dr}\end{equation}
and the term in brackets can be calculated, giving us the phase $\phi_{\Delta n}$. Finally, we can reconstruct the time-dependent change in index as:
\begin{equation}\Delta n(t)=Ae^{i\phi_{\Delta n}^{A}}e^{-t/\tau_A}+Be^{i\phi_{\Delta n}^B}e^{-t/\tau_B}+Ce^{i\phi_{\Delta n}^C}.
\end{equation}

Two \textit{caveats} are in order. Strictly, Eq. \ref{Dr} is true only at normal incidence, but as we are interested not in the amplitude of $\Delta r$ but in its complex phase, we find that oblique incidence makes only a small difference. More important is the half-angle in Eq. \ref{halfangle}, since half-angles are defined only \textit{modulo} $\pi$. To determine whether to add the $\pi$, we calculate $\Delta R(t)=|\Delta r(t)|^2$, where
\begin{equation}\Delta r(t)=Ae^{i\phi_{\Delta r}^{A}}e^{-t/\tau_A}+Be^{i\phi_{\Delta r}^B}e^{-t/\tau_B}+Ce^{i\phi_{\Delta r}^C},
\end{equation}
and we compare $\Delta R(t)$ with measurements (pump-probe experiments without the heterodyne detection), making sure that the sign and overall shape match.

\section{Signal at long times and thermal diffusion}

All of our data fit well to one of two triple-exponential functions. For the transient-grating data,
\begin{equation}\Delta r(t)=Ae^{i\theta_A}e^{-t/\tau_A}+Be^{i\theta_B}e^{-t/\tau_B}+Ce^{i\theta_C}e^{-t/\tau_C},
\label{twoexpsup}\end{equation}
while for the pump-probe data, which are real-valued,  
\begin{equation}\Delta R(t)=Ae^{-t/\tau_A}+Be^{-t/\tau_B}+Ce^{-t/\tau_C}.
\label{twoexpreal}\end{equation}
The first two terms have been discussed extensively in the main text. In this section we discuss the $C$ term, which we attribute to transient heating of the lattice. The magnitude $C$ is typically only about 20\% that of $A$ or $B$. Perhaps most important, despite adding to our fits three parameters---$C$, $\theta_C$ and $\tau_C$---the term has negligible influence on the determination of $\tau_A$ and $\tau_B$. This is because, for all the Dirac and Weyl materials studied, $\tau_C$ is much slower than $\tau_A$ and $\tau_B$, making the few-picosecond decays effectively biexponential. We measure data to 800 ps (not shown), and $\tau_C$ lies in the range of tens to hundreds of picoseconds (Table \ref{tableC}), so we obtain the parameters $C$, $\theta_C$ and $\tau_C$ unambiguously, then we hold them constant when fitting the earlier, few-picosecond decays. 

Next, we discuss measurements on two ZrSiS samples, providing evidence that this long-time signal arises from heating of the lattice. One of these samples was on a sapphire substrate, adhered with UV-curing adhesive; the other was on a Cu substrate and adhered with silver paste. The latter sample ought to be in better thermal contact with the cold-finger of the cryostat, and so should cool more rapidly after photoexcitation. The data in Fig. \ref{ThermalDiffusion} show that at each measured $q$ the decay rate of the long-time signal, $1/\tau_C$, is greater for the sample on the copper substrate. This is consistent with the hypothesis that the long-time signal originates in heat that the laser deposits in the lattice (via electronic excitation at earlier times), and that its roughly 100-ps lifetime reflects the time required for transport of heat away from the excited region into the substrate.

\begin{figure}
\includegraphics[width=2.3 in]{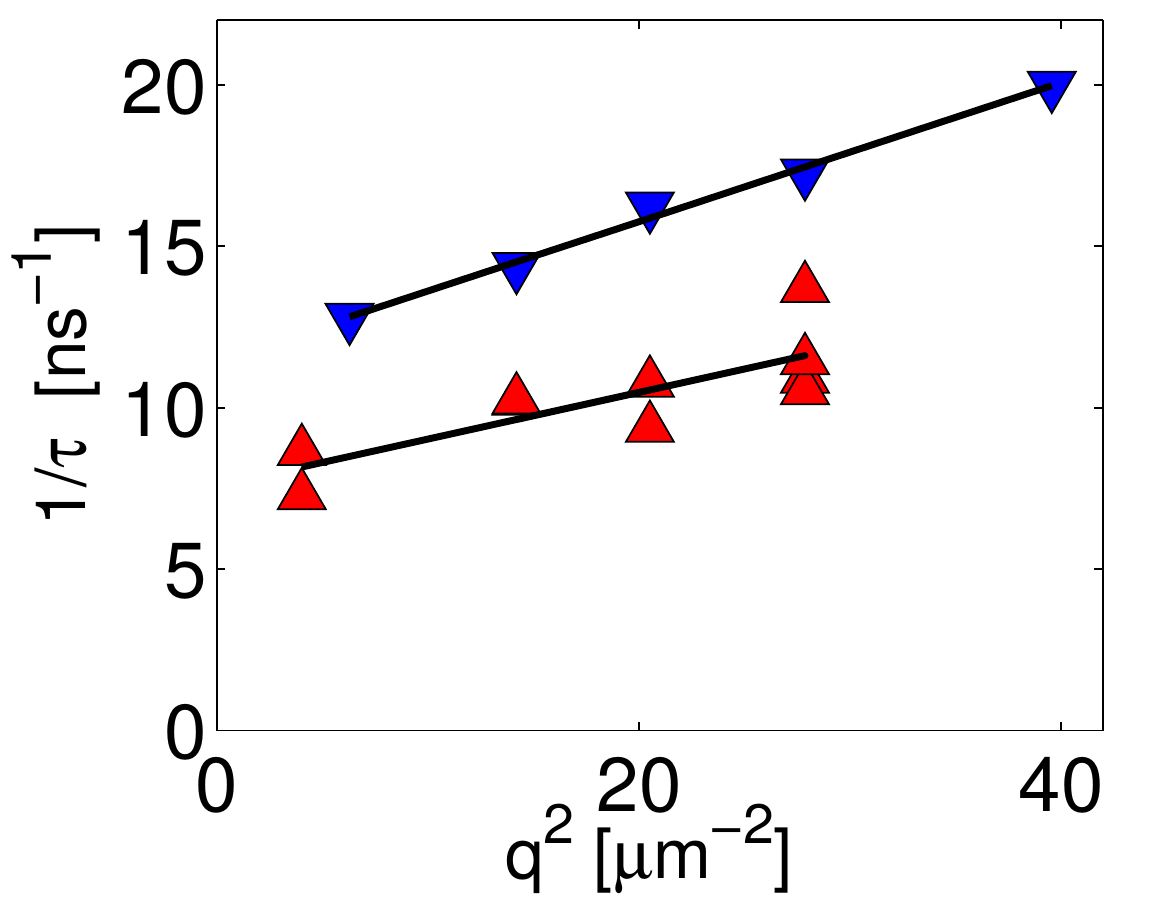}
\caption{\textbf{Slow decay in ZrSiS.} The decay rate $1/\tau_C$ of the transient grating at 8 K, as a function of the grating wavevector $q$. Up-pointing triangles are data for a ZrSiS sample on a sapphire substrate, while the downward triangles are for ZrSiS on a Cu substrate. Lines are fits to Eq. \ref{diffusioneq}.}\label{ThermalDiffusion}
\end{figure}

\begin{table}
\begin{tabular}{l|c|c|c|c}
\hline Material & SrMnSb$_2$ & TaAs & ZrSiS & NbP \\ \hline\hline
$\tau_C [ps]$ & $400-600$* & $\infty^\dagger$ &  $\tau_C>27$ & $97-190$ \\ \hline
$C/(A+B)$ & $16\%-38\%$ & $2\%-6.5\%$ &  $6\%-10\%$ & $14\%-20\%$\\ \hline
\end{tabular}
\caption{\textbf{Typical range of parameters of the slowest decay,} measured at room-temperature. (*) For SrMnSb$_2$, there is an additional decay of comparable magnitude and of timescale 13-29 ps. ($^\dagger$) For TaAs, $\tau_C$ is indistinguishable from infinity, and is fixed to infinity in fits.}
\label{tableC}
\end{table}

The $q$-dependence in Fig. \ref{ThermalDiffusion} also supports this hypothesis. Recall that in a transient-grating experiment, the ``grating'' is a spatially sinusoidal modulation of the photoinduced change in refractive index, $\Delta n(t)$, and that the measured signal is the light diffracted off of this grating. The signal may therefore decay either through the gradual relaxation of $\Delta n(t)$ or through diffusion, which reduces the contrast between the grating's peaks and its troughs. The total rate of decay due to both processes is 
\begin{equation} \frac{1}{\tau(q)}=Dq^2+\frac{1}{\tau_0},
\label{diffusioneq}
\end{equation}
where $D$ is the diffusion coefficient, and measurement at several $q$ determines $D$. As seen in Fig. \ref{ThermalDiffusion}, the decay rate $1/\tau$ follows the form of Eq. \ref{diffusioneq}, indicating diffusive motion. For the sample on sapphire, $\tau_0=127$ ps and $D=1.2$ cm$^2$/s; for the sample on Cu, $\tau_0=87$ ps and $D=2.2$ cm$^2$/s. Both of these diffusion coefficients are in the range typical of thermal diffusion, providing evidence that the source of the signal is thermal. 

\section{Biexponential decay vs. simpler functional forms}

In the main text, we have described our data using the form of a biexponential decay (Eq. 1). It would be preferable, if possible, to describe the data using a simpler function with fewer free parameters. In this section we consider two possible functions, both frequently used to describe ultrafast dynamics. We show that neither can fit our data adequately, and we discuss the reason.

The two functions we consider are single-exponential decay,
\begin{equation}\Delta r(t)=Ae^{i\theta_A}e^{-t/\tau}+Ce^{i\theta_C},
\label{singleexp}\end{equation}
and bimolecular recombination,
\begin{equation}\Delta r(t)=\frac{1}{(1/A)+\gamma(t-t_0)}e^{i\theta_A}+Ce^{i\theta_C}.
\label{bimoleculareq}\end{equation}
In these equations we have included a constant term, $C$, and the factors of $\exp(i\theta_A)$ and $\exp(i\theta_C)$ allow the phase of the decaying component to differ from that of the constant term. The variables $A$, $\theta_A$, $\tau$, $C$, $\theta_C$, $\gamma$, and $t_0$ are treated as free parameters.

\begin{figure}
\includegraphics[width=\columnwidth]{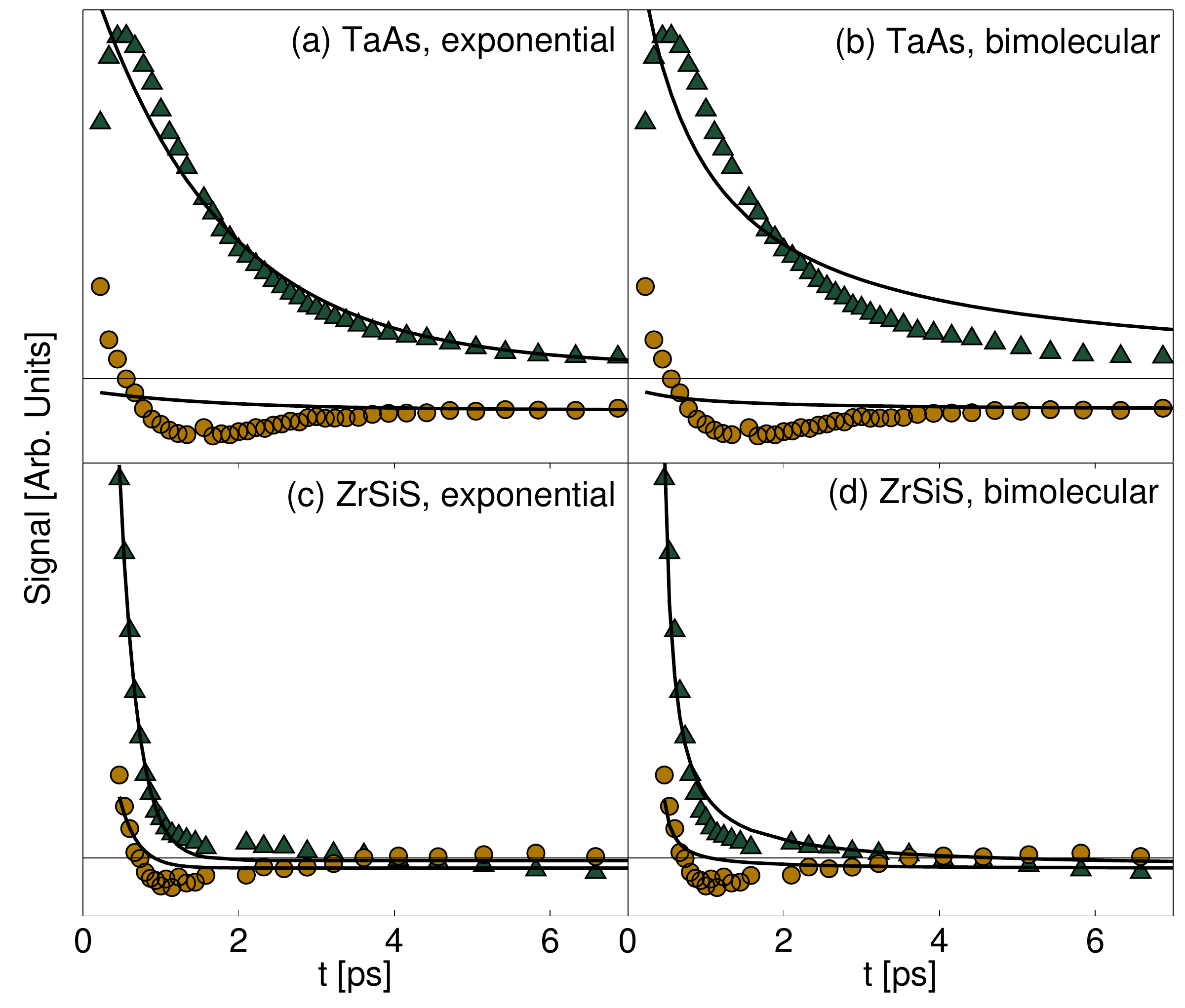}
\caption{\textbf{Alternate fits to data.} Room-temperature data for TaAs (panels (a) and (b)) and ZrSiS (panels (c) and (d)) are shown along with least-squares fits to the monoexponential form Eq. \ref{singleexp} (panels (a) and (c)) and to the bimolecular form Eq. \ref{bimoleculareq} (panels (b) and (d)). The fits are visibly worse than those to the bi-exponential form shown in the main text. Triangles and circles are real and imaginary parts, respectively, up to a factor of $e^{i\phi}$.}
\label{Bimolecular}
\end{figure}

As shown in Fig. \ref{Bimolecular}, neither of these functional forms fits our data adequately. The reason, too, is apparent from the figure: in the measured data the signal's fast ($\tau_A$) and slow ($\tau_B$) decays have different complex phases. For instance, in the real part of the signal for ZrSiS the slow decay \textit{adds} to the fast decay, while in the imaginary part the slow decay \textit{subtracts} from it. In contrast, Eq. \ref{singleexp} and Eq. \ref{bimoleculareq} each have just one decaying component, making them unable to simultaneously fit the real and imaginary parts of the measured data, no matter what value of $\theta_A$ is chosen. 

\section{Additional magnetic-field data}
\begin{figure}
\includegraphics[width=3.37 in]{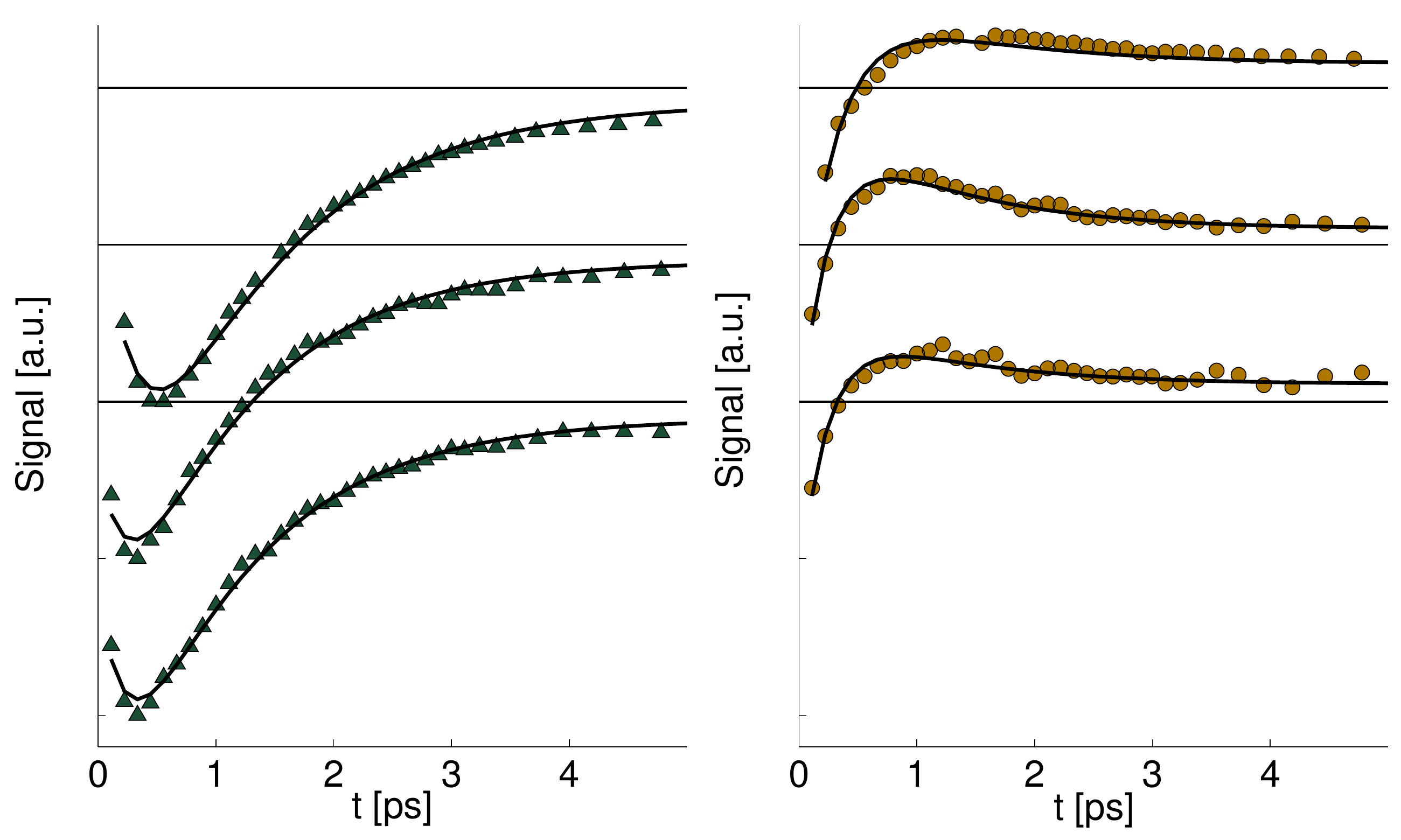}
\caption{\textbf{Transient-grating measurement of TaAs under several conditions.} Triangles and circles are real and imaginary parts, respectively (up to a factor of $e^{i\phi}$). Solid curves are fits to Eq. \ref{twoexpsup}. Curves are shifted vertically for clarity: top is room temperature, $B=0$; middle is $T=8$ K, $B=0$; bottom is $T=8$ K, $B=0.3$ T.}
\label{TaAsBfield}
\end{figure}

In Fig. \ref{TaAsBfield} we show transient-grating data from TaAs. The data compare room temperature with 8 K; at the lower temperature, data are shown both with and without a 0.3-T magnetic field. Notice that the data with an applied magnetic field look essentially the same as those without. The same is true of ZrSiS, and of Sr$_{1-y}$Mn$_{1-z}$Sb$_{2}$ (which was shown, at much higher field, in the main text).

\bibliography{Bib171031a}
\bibliographystyle{Nature}

\end{document}